\documentclass[12pt]{article}
\usepackage{latexsym}
\def\ab{C_{AB}}
\def\ba{C_{BA}}

\begin{document}

\title{Topological mass mechanism and exact fields mapping}
\author{R. L. P. G. Amaral$^a$, O.S. Ventura$^{b,c}$, \\ L. O. Buffon$^{c,d}$ and J V Costa$^{a,e}$, 
\footnote{ rubens@if.uff.br, ozemar@uvv.br, costa@if.uff.br, lobuffon@terra.com.br }\\
\small\em $^a$Instituto de F\'\i sica, Universidade Federal Fluminense, 24210 - 340, Niter\'oi - RJ, Brasil,\\
\small\em $^b$Coordenadoria de Matem\'{a}tica, Centro Federal de Educa\c {c}\~{a}o Tecnol\'{o}gica \  do Esp\'{\i}rito Santo,  \\
\small\em $ $Avenida Vit\'{o}ria 1729 - Jucutuquara, Vit\'{o}ria - ES, 29040 - 333, Brazil\\
\small\em $^c$Centro Universit\'{a}rio de Vila Velha, Rua comiss\'{a}rio Jos\'{e} \\
\small\em Dantas de Mello 15 - Boa Vista, Vila Velha - ES, 29102 - 770, Brazil\\
\small\em $^d$Escola Superior de Ci\^{e}ncias da Santa Casa de Miseric\'{o}rdia de Vit\'{o}ria,\\ \small\em Av. Nossa Senhora da Penha 2190, Santa Luiza, Vit\'{o}ria-ES, 29045-402, Brazil\\
\small\em $^e$FAESA, Campus I, Rua Anselmo Serrat,199, Ilha Monte Belo, Vit\'oria-ES, 29040-410, Brazil\\}

\date{\today}

\bigskip
\maketitle

\begin{abstract}

We present a class of mappings between models with topological mass mechanism 
and purely topological models in arbitrary dimensions. These mappings are established 
by directly mapping the fields of one model in terms of the fields of the other model in
 closed expressions. These expressions provide the mappings of their actions as well as the
 mappings of their
propagators. For a general class of models in which the topological model becomes
the BF model the mappings present arbitrary functions which otherwise are absent for Chern-Simons like actions. This work generalizes the results of \cite{ozemar} for arbitrary dimensions.

\setcounter{page}{0}\thispagestyle{empty}

\end{abstract}

\section{Introduction}

The search for ultraviolet renormalizable models has always been one of the
most attractive and relevant aspects of quantum field theory. As it is well
known, the program of describing the electro-weak interactions in the language
of QFT is based on the construction of the Higgs mechanism
for mass generation of the vector bosons. However, this mechanism relies on
the existence of a scalar particle, the Higgs boson, whose experimental
evidence is still lacking.

In this context, the topological mechanism for mass generation is attractive,
since it provides  masses for the gauge vector bosons without the explicit
introduction of new scalar fields. For example, in three-dimensional spacetime, the
topological non-Abelian Chern-Simons term generates mass for the Yang-Mills
fields while preserving the exact gauge invariance $\cite{ng1}$. 
Indeed an important property of the three dimensional Yang-Mills type actions, in
the presence of the Chern-Simons term, was pointed out in  $\cite{simple}$,
i.e., it can be cast in the form of a purely Chern-Simons action through a
nonlinear covariant redefinition of the gauge connection  $\cite{ng3}$. The quantum
consequences of this fact were investigated in the BRST framework yielding
an algebraic proof of the finiteness of the Yang-Mills action with
topological mass $\cite{setup}$.

In four dimensions
the topological  mass generation mechanism occurs in the case of an anti-symmetric 
tensorial
field $B_{\mu \nu }$. It has been shown that the Cremmer-Sherk  action
gives a massive pole to the vector gauge field in the Abelian context $\cite{ng2}$.
Indeed, as it was shown $\cite{nogo},$ this model exists only in the Abelian
version. In fact, possible non-Abelian generalizations of the Cremmer-
Sherk action  will necessarily require non-renormalizable couplings,
as in $\cite{ng4}$, or the introduction of extra fields $\cite{ng5}$.
Anti-symmetric rank-two fields in four dimensions also deserve attention since they
appear naturally by integrating out the fermionic degrees of freedom  in favor
of bosonic fields in bosonization approaches. The fermionic current turns
out to be expressed in terms of derivatives of a tensorial field as a topologically
conserved current quite similar to the expression of the current in terms of vectorial fields in
three dimensions \cite{marino1}. The coupling of this current 
to the gauge field leads to terms in the effective action similar to the one of the
Cremmer-Sherk model \cite{botta}.
The mapping of the Cremmer-Sherk's
action to the pure BF action, in an iterative way, was presented in \cite{Landim} both in the Abelian as in the non-Abelian cases. The exact forms of these mappings in the four as well as in the three dimensional
cases were presented in $\cite{ozemar}$.

In D dimensions
the topological  mass generation mechanism occurs in a class of models with an anti-symmetric 
p-form field $B$ coupled to a q-form field $A$, with $q+p=D-1$.  The interest for these models
also appears from the bosonization procedure in arbitrary dimensions which is known to lead to
antisymmetric tensorial fields with higher ranks \cite{botta}.

In this work we generalize the results shown in $\cite{ozemar}$  by presenting the exact
mapping between generalized Cremmer-Sherk's actions
in arbitrary dimensions and purely topological $BF$ models. The closed expression we obtain
depends on arbitrary functions and has a  non-local nature.
We also obtain the mapping from Maxwell-Chern-Simons like models and pure Chern-Simons
ones  in higher dimensions.

\section{Generalized Cremmer-Cherk Actions}
\setcounter{equation}0
The aim of this section is to establish the classical equivalence between the
generalized Cremmer-Sherk's action and the pure $BF\,$ theory, \textit{i. e.}, that the
first action can be mapped to the second one through a redefinition of
the antisymmetric fields. Following the same steps of the four-dimensional case $\cite
{ozemar}$, we search for a  redefinition 
 of the q-form gauge field $A $ and the p-form gauge field $B$, subject to $q+p=D-1$, as  closed expressions in terms
of corresponding field   $\hat A $ and $\hat B$ in
 such a way that the equivalence of their actions can be asserted\footnote{  We work in  the Minkovski spacetime with $\eta_{\mu\nu}=\mbox{Diag}(+,-,...,-)$, $\varepsilon^{123...D}=1$  and $\varepsilon^{\mu_1\mu_2..\mu_D}
=\eta^{\mu_1\nu_1}...\eta^{\mu_D\nu_D}\,\varepsilon_{\nu_1\nu_2..\nu_D}$}: 
\begin{equation}
{\mathcal{S}_M}(A;N_A)\;+\;{\mathcal{S}_H}(B;N_B)+{\mathcal{S}_{BF}}(A,B;N_{AB})=
{\mathcal{S}_{BF}}(\widehat{A},\widehat{B};\widehat N_{AB})\;.  \label{bff1}
\end{equation}

Here 
\begin{eqnarray}\label{csandbf}
{\mathcal{S}_M}(A;N_A) &=&\int  \;A\wedge d*d N_A A     
\,, \\
{\mathcal{S}_H}(B;N_B) &=&\int  \;B\wedge d*d N_B B   
 \,, \\\label{purebf}
{\mathcal{S}_{BF}}(A,B;N_{AB}) &=&\int  \;B\wedge d N_{AB} A    \,,
\end{eqnarray}
\noindent where $N_A$, $N_B$, $N_{AB}$ and $\widehat N_{AB}$ are generic scalar operators
which commute with the operations $*$ and $d$. In the simplest case they will be
 just normalization factors.
Indeed taking the field redefinitions in the form

\begin{eqnarray}\label{mapping1}
 A&=& \widehat A+  C_{AA}\,*d*d\,\widehat A+ C_{AB}*d\,\widehat B ,\nonumber\\
 B&=&\widehat B+ C_{BB}\,*d*d\,\widehat B+  C_{BA}*d\,\widehat A
\end{eqnarray}

\noindent and using this mapping in the equation (\ref{bff1}) the following set of equations are obtained:

\begin{eqnarray}\label{crsh}
N_BC_B^2+ \ab^2N_A \Delta
-N_{AB}\ab C_B &=&0,\nonumber\\
N_B\ba^2\Delta 
+N_AC_A ^2-N_AC_A \ba&=&0,\nonumber\\
2N_B\ba C_B  \Delta +\,2N_A C_A  \ab
 \Delta +\;\;\;\;\;\;\;&&\nonumber\\
{}\;\;\;\;\;\;\;\;\;\;\;\;\;\;\;\;- N_{AB}\ab\ba \Delta 
-N_{AB}C_A C_B 
-\widehat N_{AB}&=&0.
\end{eqnarray}

\noindent We have defined $C_{A}=1+
 C_{AA}*d*d$ and $C_{B}=1+
 C_{BB}*d*d$. 
Let us mention that the action of the Delta operator on a p-form ${ w}$ is given by
$$\Delta {w}=(-1)^{Dp}(d*d*+(-1)^D*d*d)w=-g^{\mu\nu}\partial_{\mu}\partial_{\nu}w.$$

\noindent 
Note   that $\Delta$ commutes with the operations $*$ and $d$. Also  for forms
$w$  and $v$ 
with compact support it follows that

\begin{equation}\label{integralbyparts}
\int w\wedge\Delta v=\int\Delta w\wedge v.
\end{equation}
\noindent This means that the operator $\Delta$ can be dealt with as if it were an scalar.
The same properties will be warranted to the coefficients $C_{XY}$ (with $X$ and $Y$ meaning 
$A$ and/or $B$).

The coefficients that solve eqs. (\ref{crsh}) turn out to be 

\begin{eqnarray}\label{strucfunc}
 C_{AA}&=&\frac{(-1)^{D(D-p)}\widehat N_{AB}^{1/2}}{ \Delta }\left[{\frac{2^{1/2}}{ \sigma} }\left\{ \frac{N_{AB}+\varepsilon \left[
N_{AB}^2-4N_AN_B \Delta
\right] ^{1/2}}{
\left( N_{AB}^2-4N_AN_B\Delta
 \right) }\right\} ^{1/2}-\frac 1{N_{AB}^{1/2}}\right],\nonumber\\
 C_{BB}&=&\frac{(-1)^{D(p+1)}\widehat N_{AB}^{1/2}}{ \Delta }\left[\frac{ \sigma}{2^{3/2} }\left\{ \frac{N_{AB}+\varepsilon \left[
N_{AB}^2-4N_AN_B \Delta
\right] ^{1/2}}{
\left( N_{AB}^2-4N_AN_B\Delta
 \right) }\right\} ^{1/2}-\frac 1{N_{AB}^{1/2}}\right],\nonumber\\
 C_{AB}&=&\frac{\sigma \widehat N_{AB}^{1/2}}{2^{3/2}N_A \Delta}
\left\{ \frac{N_AN_B \Delta 
 \left( N_{AB}-\varepsilon \left[ N_{AB}^2-4N_AN_B \Delta \right] ^{1/2}\right) }{\left(
N_{AB}^2-4N_AN_B \Delta
\right) }\right\} ^{1/2}
 ,\nonumber\\
 C_{BA}&=&\frac{2^{1/2} \widehat N_{AB}^{1/2}}{ \Delta\sigma N_B}
\left\{ \frac{N_AN_B \Delta 
 \left( N_{AB}-\varepsilon \left[ N_{AB}^2-4N_AN_B \Delta \right] ^{1/2}\right) }{\left(
N_{AB}^2-4N_AN_B \Delta
\right) }\right\} ^{1/2}.  
\end{eqnarray}
\noindent Here $\sigma$ are operators we discuss in the following while $\epsilon$ is inserted
just to keep control of the branches of the square root.

Looking for the inverse mapping we search for $\widehat C_{XY}$ such that
\begin{eqnarray}\label{mapping1inv}
\widehat A&=&  A+ \widehat C_{AA}\,*d*d\, A+\widehat C_{AB}*d\,B ,\nonumber\\
\widehat B&=& B+\widehat C_{BB}\,*d*d\,B+ \widehat C_{BA}*d\,A.
\end{eqnarray}

\noindent Using (\ref{mapping1}) in (\ref{mapping1inv}), we find after some manipulations  

\begin{eqnarray}\label{strucfunc2}
\widehat C_{AA}&=&\frac{(-1)^{D(D-p)}}{\widehat N_{AB}^{1/2} \Delta }
 \left[{  \frac {2^{1/2}} {\widehat \sigma}   }  
 \left\{ {N_{AB}+\varepsilon \left[
N_{AB}^2-4N_AN_B \Delta
\right] ^{1/2}}    \right\} ^{1/2}-N_{AB}^{1/2}\right]
,\nonumber\\
 \widehat C_{BB}&=&\frac{(-1)^{D(p+1)}}{\widehat N_{AB}^{1/2} \Delta }\left[\frac{N_A\widehat\sigma }{ 2^{1/2}}\left\{   {N_{AB}+\varepsilon \left[
N_{AB}^2-4N_AN_B \Delta
\right] ^{1/2}}   \right\} ^{-1/2}-N_{AB}^{1/2}\right],\nonumber\\
\widehat C_{AB}&=&\frac{N_B 2^{3/2}}{\widehat \sigma
\widehat N_{AB}^{1/2}}\left\{   {N_{AB}+\varepsilon \left[
N_{AB}^2-4N_AN_B \Delta
\right] ^{1/2}}   \right\} ^{-1/2} ,\nonumber\\
\widehat C_{BA}&=&\frac{ \widehat \sigma }{2^{3/2}
\widehat N_{AB}^{1/2}}\left\{   {N_{AB}+\varepsilon \left[
N_{AB}^2-4N_AN_B \Delta
\right] ^{1/2}}   \right\} ^{1/2}.  
\end{eqnarray}

\noindent Observe  that the presence of the arbitrary operator $\sigma$ or $\widehat\sigma$ in the 
equations (\ref{strucfunc}) and (\ref{strucfunc2})  has been 
noted in the four dimensional case \cite{ozemar}. It is 
expected since the set of transformations

\begin{eqnarray}
\hat A &\longrightarrow& \sigma \hat A,\nonumber \\
\hat B &\longrightarrow& \frac{1}{\sigma}\hat B,
\end{eqnarray}
does not affect the generalized BF model action.

Let us now  investigate  the usual Cremmer-Sherk model, that is, the case in which the
arbitrary operators present in the actions are reduced to normalization factors that can be fixed as
\begin{eqnarray}
N_{AB}&=& (-1)^{p}\,m,\\
N_{A}&=&N_B=\frac{(-1)^{D(p+1)+1}}{2}.
\end{eqnarray}

\noindent In this case the mapping is defined by

\begin{eqnarray}\label{csusual}
C_{AA}&=&\frac{4(-1)^{D^2-1}}{\sigma^2} C_{BB}=\frac{(-1)^{D(D-p)}}{ \Delta }\left[{\frac{(2m)^{1/2}}{ \sigma} \left[\frac{m\pm \sqrt{m^2-\Delta}}{m^2-\Delta}
\right]^{1/2}}-1\right],\nonumber\\
C_{AB}&=&\frac{\sigma^2} 4C_{BA}=\frac{\sigma}4\sqrt{\frac{2m}{\Delta}}\left[\frac{m\mp\sqrt{m^2-\Delta}}{m^2-\Delta}\right]
^{1/2}
\end{eqnarray}

\noindent This case has been dealt with in \cite{Landim}  within an approach whereby the
coefficients are defined as a series whose terms are obtained iteratively after 
fixing arbitrarily the freedom in the mapping which we have explained above.
A generalization of the iterative mappings provided in that work may be retrieved from our 
procedure  defining
conveniently the operator $\sigma$(or $\hat \sigma$). By expressing

\begin{equation}
\sigma=\sum_{n=0}^\infty C_n\left(\frac{ \Delta N_A N_B} {N_{AB}^2}\right)^n , \nonumber \\
\end{equation}

\noindent and expanding the exact mapping (\ref{csusual}) in powers of $\Delta$ a
 series of increasing  powers
of $ \Delta$ is easily obtained. In this setting the operator $\sigma$ will give rise to a new independent parameter $C_n$ at each new order of the series expansion. Indeed the 
series obtained iteratively in \cite{Landim} corresponds to a particular choice of such set 
of constants.

On the other hand the operators defining the mapping can be alternatively expanded
in powers of $N_{AB}$ instead of powers of $1/N_{AB}$. Let us first consider the particularly important case corresponding to the limit $N_{AB}\longrightarrow 0$
in equations (\ref{strucfunc2}). This leads to the mapping from the fields of a model without topological
terms to the purely topological model fields: 

\begin{eqnarray}\label{strucfunc2b}
\widehat C_{AA}&=&\frac{2(-1)^{D(D-p)+1}\sqrt\epsilon [-N_AN_B\Delta]^{1/4}}{\hat N_AB^{1/2}\Delta\hat\sigma},
\nonumber\\
\widehat C_{BB}&=&\frac{(-1)^{D(D-p)+1)} N_A\hat\sigma}{2\hat N_{AB}^{1/2}\Delta[-N_AN_B\Delta]^{1/4}},\nonumber\\
\widehat C_{AB}&=&\frac {2N_B}{\hat\sigma\hat N_{AB}^{1/2}\sqrt{\epsilon}[-N_AN_B\Delta]^{1/4}} , \nonumber\\
\widehat C_{BA}&=&\frac {\hat \sigma\sqrt\epsilon [-N_AN_B\Delta]^{1/4}}{2\hat N_{AB}^{1/2}}.  
\end{eqnarray}
 
 The inverse of this mapping may be directly obtained. This inverse mapping can also be obtained performing 
the limit $N_{AB}\longrightarrow 0$ already in the structure functions (\ref{strucfunc}).
A subtle technical point is that the main contribution that goes with $1/ \Delta$ should be
eliminated by first reabsorbing in equations (\ref{mapping1})  the first terms of the right hand side of those equations. This
 amounts to a change in the gauge fixing conditions. 
 
 This limiting case can also be interpreted as
 the zeroth order term of  the alternative expressions in powers of $N_{AB}$ of the coefficients given in (\ref{strucfunc2}).
 Indeed the direct alternative power series  expansion of (\ref{strucfunc2}) turn out to result in a set of series 
in powers of $N_{AB}/\sqrt{ -\Delta}$ multiplying its zeroth order expressions (\ref{strucfunc2b}).
These series are essencially the same ones that have been made explicit in \cite{ozemar} in the four dimensional case. 
 
 These series can be alternatively obtained in a procedure that parallels the one used to 
obtain the iterative mapping in powers of $ \Delta/N_{AB}^2$. For this (\ref{strucfunc2b}) should be 
taken as the zeroth order
expression, that maps the action ${\mathcal{S}_M}(A)\;+\;{\mathcal{S}_H}(B)$ to $
{\mathcal{S}_{BF}}(\hat A,\hat B)$. Now the perturbation $
{\mathcal{S}_{BF}}(A,B)$ is taken into account and its contribution is canceled at each step of
the process with higher order terms. Comparing to the previous procedures the roles of the kinetic 
terms are thus reversed. The presence of $\sqrt{- \Delta}$ in these series may seen suspicious at first 
sight. After all if the propagators of the Cremmer-Sherk model fields are expanded in powers
of $N_{AB}$ they turn out to produce, instead, a series of powers of $N_{AB}^2/ \Delta$. Indeed an explicit computation shows that
when the series obtained by expanding (\ref{strucfunc}) is used to obtain the propagators of
$A$ and $B$ from the ones of  $\hat A$ and $\hat B$ the terms with square roots of the D'Alembertian
cancel out.

Note that the gauge symmetry of the  actions can be expressed
using the (n-p-2)-form $a$ and the (p-1)-form $b$ as

\begin{equation}
\delta ^g A =d\, a
,\,\,\,\,\,\,\,\,\,\,\,\,\,\,\,\delta ^gB =0  \label{bhg1}
\end{equation}
and 
\begin{equation}
\delta ^tA  =0,\;\;\;\;\;\,\,\delta ^tB = d\,b.  \label{mjk1}
\end{equation}

\noindent while for the topological model fields we have the symmetries
\begin{equation}
\delta ^g \widehat A =d\, \widehat a
,\,\,\,\,\,\,\,\,\,\,\,\,\,\,\,\delta ^g\widehat B =0  \label{bhg2}
\end{equation}
and 
\begin{equation}
\delta ^t\widehat A  =0,\;\;\;\;\;\,\,\delta ^t\widehat B = d\,\widehat b.  \label{mjk2}
\end{equation}

The mapping (\ref{mapping1}) has been chosen in such a way that
one pair of fields differs from the other  in terms of gauge invariant terms. 
With this the gauge transformations of one pair of fields is translated
straightforwardly into the ones of the other pair. That is, this leads to the identification
of $\widehat a$ and
$\widehat b$ with $a$ and $b$ respectively. Of course other choices of mappings can be done
which would result in different relations among the gauge parameters of each model but which still map their actions and thereby their free propagators.

\section{Chern-Simons Field}

Let us consider now, in $D=4n-1$, the generalized Chern-Simons action given
by

\begin{equation}
S_{CS}=\int   A\wedge \hat{\cal D}dA.
\end{equation}

\noindent Consider also the generalized Maxwell-Chern-Simons action

\begin{equation}
S_{MCS}= \int  \left(  A\wedge d {\cal D}\,A +  F{\cal C}\wedge*F\right) .
\end{equation}

\noindent Here the factors  ${\cal C}$ and ${\cal D}$ are generic scalar operators that obey the same rule as  $\Delta$  in eq. (\ref{integralbyparts}).

We are searching for a mapping from the field $\hat A$ with pure Chern-Simons action to the Maxwell-Chern-Simons field $A$ in the form

\begin{equation}\label{csmapping}
\hat A= -f*dA+g*d*dA.
\end{equation}

\noindent Using this expression in the pure Chern-Simons action and requiring the outcome of the Maxwell-Chern-Simons action gives the set of equations

\begin{eqnarray}
-fg\Delta&=&\frac {\cal C}{2\hat{\cal D}},\nonumber\\
f^2\Delta +g^2\Delta^2&=&\frac {\cal D}{\hat{\cal D}}.
\end{eqnarray}

\noindent The mapping becomes then

\begin{eqnarray}\label{cscoefficients}
f&=& -\left(\frac{\cal C}{2\hat{\cal D}}\right)^{1/2}\left[\frac{\cal D}{\cal C}+\epsilon\left(\frac {{\cal D}^2}{{\cal C}^2}-\Delta\right)^{1/2}\right]^{-1/2},\nonumber \\
g&=&\left(\frac{\cal C}{2\hat{\cal D}}\right)^{1/2}\frac{1}{\Delta}\left[\frac{\cal D}{\cal C}+\epsilon\left(\frac {{\cal D}^2}{{\cal C}^2}-\Delta\right)^{1/2}\right]^{1/2},
\end{eqnarray}
\noindent where $\epsilon=\pm 1$ controls the different branches of the square root.

The mapping can be inverted by setting

\begin{equation}\label{csmappinginverted}
 A= -\hat f*d\hat A+\hat g*d*d\hat A
\end{equation}
and using this expression in (\ref{csmapping}) and (\ref{cscoefficients}). The new structure functions 
turn out to be given by

\begin{eqnarray}
\hat f g+\hat gf&=&0,\nonumber\\
\hat gg\Delta^2+\hat ff\Delta&=&1.
\end{eqnarray}
The solution gives the inverse mapping coefficients as

\begin{eqnarray}\label{cscoefficientsinverted}
 \hat f&=& \epsilon\left(\frac{\hat{{\cal D}}}{2{\cal C}}\right)^{1/2}\left[
 \frac{{
 \frac{\cal D}{\cal C}}-\epsilon\left(\frac{{\cal D}^2}{{\cal C}^2}-\Delta\right)^{1/2}}
 {\Delta\left(
 \frac{{\cal D}^2}{{\cal C}^2}-\Delta\right)}\right]^{1/2},\nonumber\\
 \hat g&=& -\epsilon\left(\frac{
  \hat{{\cal D}}}{2{\cal C}}\right)^{1/2}\frac{1}{\Delta}\left[
 \frac{{
 \frac{\cal D}{\cal C}}+\epsilon\left(\frac{{\cal D}^2}{{\cal C}^2}-\Delta\right)^{1/2}}
 {
 \frac{{\cal D}^2}{{\cal C}^2}-\Delta}\right]^{1/2}.
 \end{eqnarray}

 Let us consider some limiting cases of interest. The case where ${\cal C}=1/2 ={\cal D}/m$ corresponds to the mapping from the usual
  Maxwell-Chern-Simons model to the Chern-Simons model. This case in $D=3$ has been dealt with in \cite{ozemar}. Secondly let us consider the limit where ${\cal D}\longrightarrow 0$. When ${\cal C}=\hat{{\cal D}}=1/2$   this limit corresponds  to the pure Maxwell field action. In that case we have

\begin{eqnarray}\label{struturem}
\hat g&=&\sqrt{\frac{\hat {\cal D}}{2{\cal C}}}(-\Delta)^{-\frac{5}{4}}\nonumber\\
\hat f&=&\sqrt{\frac{\hat {\cal D}}{2{\cal C}}}(-\Delta)^{-\frac{3}{4}}.
\end{eqnarray}

\noindent Another limit of particular interest is given by taking ${\cal C}\longrightarrow 0$, while ${\cal D}=\hat{{\cal D}}=1/2$. This means to perform the limit from pure Chern-Simons to pure Chern-Simons
models. The trivial mapping $\hat A\equiv A$ due to gauge freedom can be expressed in the setting of the expression (\ref{csmappinginverted}) with

\begin{eqnarray}
\hat g&=&1/\Delta\nonumber\\
\hat f&=&0.
\end{eqnarray}

\noindent It is interesting to point out that besides this trivial mapping, by choosing conveniently the square root signals,
the alternative mapping given by $\hat g=0$ and 
\begin{equation}
\hat f=\left(\frac{-\hat{{\cal D}}}{\cal D}\right)^{1/2}\left(\frac{1}{-\Delta}\right)^{1/2}
\end{equation}
can be obtained.

The two limiting cases above may be considered as the zeroth order approximation of the
two alternative iterative representations of the mappings. Let us first present the general
form for the expansion in powers of $(-\Delta{{\cal C}}^2/{{\cal D}}^2)$.

\begin{eqnarray}\label{g}
\hat g&=&\frac1{\Delta}\left[1+\frac{3\Delta{{\cal C}}^2}{2{{\cal D}}^2}+\frac{35\Delta^2{{\cal C}}^4}{128{{\cal D}}^4}+\frac{231\Delta^3{{\cal C}}^6}{1024
{{\cal D}}^6}+\frac{6435\Delta^4{{\cal C}}^8}{32960{{\cal D}}^8}+....\right]\\ \label{f}
\hat f&=&\frac{-{\cal C }}{2{\cal D}}\left[1+\frac{5\Delta{{\cal C}}^2}{2^3{{\cal  D}}^2}+\frac{63\Delta^2{{\cal C}}^4}{2^7{{\cal D}}^4}+\frac{
429\Delta^3{{\cal C}}^6}{2^{10}{{\cal D}}^{6}}+....\right] .
\end{eqnarray}

\noindent This series has been obtained in \cite{simple} in the $D=3$ case through an iterative procedure.
  The exact mapping displays the non-local feature of the structure functions which may be somewhat masked in the direct iterative procedure.

Next let us present the  series expansion in powers of ${\cal D}/\sqrt{-\Delta}{\cal C}$:

 \begin{equation}\label{gv} 
\hat g=\sqrt{\frac{\hat {\cal D}}{2{\cal C}}}\left(-\Delta\right)^{-\frac{5}{4}}\left[1+\frac{1}{2}\frac{{\cal 
D}}{{\cal C}\sqrt{-\Delta}}    -\frac{3}{8} \left(\frac{{\cal 
D}}{{\cal C}\sqrt{-\Delta}}  \right)^2-\frac{5}{16} \left(\frac{{\cal 
D}}{{\cal C}\sqrt{-\Delta}}   \right)^3+...\right]
\end{equation}
and 
\begin{equation}\label{fv} 
\hat f=\sqrt{\frac{\hat {\cal D}}{2{\cal C}}}\left(-\Delta\right)^{-\frac{3}{4}}\left[1-\frac{1}{2}\frac{{\cal 
D}}{{\cal C}\sqrt{-\Delta}}  -\frac{3}{8} \left(\frac{{\cal 
D}}{{\cal C}\sqrt{-\Delta}} 
 \right)^2+\frac{5}{16} \left(\frac{{\cal 
D}}{{\cal C}\sqrt{-\Delta}}   \right)^3+...\right] .
\end{equation}

This series expression can be derived starting from  the zeroth order mapping
pointed out above from the pure Maxwell to the pure Chern-Simons model and incorporating the Chern-Simons term iteratively.

Notice the interesting property that  although the mapping of the field involves odd and even
powers of the parameter ${\cal C}/{\cal D}\sqrt{-\Delta}$ when computing the correlation functions for the $A$ field using the series expansion
only even powers of the parameter persist. Indeed this comes out since the series for
$\hat f$ and $\hat g$ differ, essentially, for the signals of the odd order terms. 

\section{Conclusions and Discussions}

In this work we have studied the procedure that allows to map the Maxwell-Chern-Simons field
to the pure Chern-Simons field in  (4n-1)D
and the Cremmer-Sherk model to the  Abelian
version of the BF model in arbitrary dimensions.
An striking importance of these mappings stems from the fact that the 
dynamical mass mechanism, which occurs in the topologically massive models,
is allowed to be described in the context of purely topological models.
The latter models presenting physical contents remarkably distinct from the
former ones may offer new insights on this physical mechanism.

   It is also remarkable that the mappings above presented allow for obtaining the time-ordered functions of one model in terms of the time-ordered functions of the other model. 
In this sense the mapping establishes not only a classical relation among the models but also a relation in the
quantum sense. The possibility of obtaining the Green functions of the topologically massive
models from the ones of the topological models which present scale invariance
 may offer valuable computational advantages. Nevertheless it is important to keep in
mind that the possibility of obtaining the correlation functions of one model from
a non-local mapping of the correlation functions of the other model
does not imply the equivalence of the Hilbert space of both models.

 The BF mapping has been established within
an exact general procedure. One remarkable aspect 
that emerges  is the presence of a great deal of freedom in this
mapping.  This freedom has been elucidated as in the four dimensional case \cite{ozemar}
as due to the form of the purely topological action which is defined
through mixed products of fields. The  invariance under rescaling of 
the fields of the
BF type action is responsible for it. 

Although the topological terms are not required to establish the mappings we
discuss here the topologically massive cases allows for natural series expansions of the exact mappings. Indeed we presented two forms of series expansions which typically
can be classified as infrared and ultraviolet series.
The knowledge of the exact mappings provides us with a typical scale,
given by the topological mass parameter. The two types of mapping shown here can be
 used for instance for computing loop variables of the generalized Cremmer-
Sherk (or Maxwell-Chern-Simons)  model using the corresponding expressions of the pure BF (or Chern-Simons) models. This suggests to perform the
computation in closed fashion without resource to expansions given
by the iterative mapping.  In any case the mass parameter may provide
valuable hints to discern in which cases computations using the usual iterative
mapping \cite{sorella1, Landim} should or not be considered reliable. It can even provide alternative
expansions for instance in direct powers of the mass parameter instead of the inverse 
power series. Let us also comment that it is to be expected \cite{setup} that the
 introduction of arbitrary gauge invariant interaction terms can be absorbed
 by considering nonlinear mappings.

In order to  properly appreciate the physical meaning of the mapping, it
is important 
 to call attention to the necessity of defining the physical content
of a local field theory in terms of the local polynomial algebra of 
observable fields. The mapping here provided relates two local models each with
its physical Hilbert space reconstructed from the Wightmann functions of its own polynomial
algebra \cite{strocchi,belvedereerubens}. Since the mapping involves  non-local functions
it should be clear that within the pure BF model there are two Hilbert spaces to be obtained.
One Hilbert space is obtained from the  local polynomial algebra of
fields defined after expressing the Cremmer-Sherk fields non-locally in terms of the pure BF model
fields and it should not be confused
with the other one, the Hilbert space of the pure BF model itself. This later is obtained from its local polynomial 
algebra of fields. Although constructed with the same model fields the first Hilbert space is not  isomorphic to the second one. Instead, it will be isomorphic to the
 Hilbert space of the Cremmer-Sherk model. The same reasoning goes in the other
 direction of the mapping. In this context it is clear that neither
 Hilbert space should be considered as a subspace of the other. It is not a mapping of
 physical states that is being addressed here but a non-local mapping among the fields.

\section{Acknowledgments}

We  acknowledge Faperj, FUNCEFETES and UVV for partial financial support.


\vspace{1.0in}

\begin{thebibliography}{9}

\bibitem{ozemar}O. S. Ventura, R. L. P. G. Amaral, J. V. Costa, L. O. Buffon and V.E. R. Lemes,   J. Phys. {\bf A: Math. Gen. 37}  (2004) 11711-11723.
 
\bibitem{ng1}  R. Jackiw and S. Templeton, Phys. Rev. {\bf D23}(1981)2291;
\\W. Siegel,\ Nucl. Phys. {\bf B156}(1979)135;\\J.F. Schonfeld, 
Nucl. Phys. {\bf B185}(1981)157;\\S. Deser, R. Jackiw and S. Templeton, 
Ann. Phys. (NY) {\bf 140}(1982)372.

\bibitem{simple}  V.E.R. Lemes, C. Linhares de Jesus, C.A.G. Sasaki, S.P.
Sorella, L.C.Q.Vilar and O.S. Ventura, Phys. Lett. {\bf B418} (1998) 324.

\bibitem{ng3}  G. Barnich and M. Henneaux, Phys. Lett. {\bf B311}(1993)123.

\bibitem{setup}  V.E.R. Lemes, C. Linhares de Jesus, S.P.Sorella, L.C.Q.
Vilar and O.S.Ventura, Phys. Rev. {\bf D58} (1998) 045010;

\bibitem{ng2}  E. Cremmer and J. Scherk, Nucl. Phys. {\bf B72}(1974)117;
\\C.R. Hagen, Phys. Rev. {\bf D19}(1979)2367;\\T.J. Allen, M.J. Bowick
and A. Lahiri, Mod. Phys. Lett. {\bf A6}(1991)559;\\R. Amorim and J.
Barcelos-Neto, Mod. Phys. Lett. {\bf A10}(1995)917.

\bibitem{nogo}  L.C.Q.Vilar, V.E.R. Lemes, O.S. Ventura, C. Sasaki and S.P.
Sorella, Phys. Lett. {\bf B410} (1997) 195.

\bibitem{ng4}  D.Z. Freedman and P.K. Towsend, Nucl. Phys. {\bf B177}
(1981)282.

\bibitem{ng5}  A.Lahiri, Generating Vector Boson Masses, Phys. Rev.
{\bf 55}(1997)5045;\\D.S. Hwang and C.Y. Lee, J. Math Phys. {\bf 38}(1997)30;%
\\J. Barcelos-Neto, A. Cabo and M.B.D. Silva, Z. Phys. {\bf C72}(1996)345%
.

\bibitem{marino1} E. C. Marino, Phys. Lett.  {\bf B263} 63 (1991); D. Barci,
 C. D. Fosco and L. E. Oxman, Phys. Lett.  {\bf B375} 267 (1996); R. 
Banerjee and E. C. Marino, Nucl. Phys.  {\bf B507} 501 (1997).


\bibitem{botta} M. Botta Cantcheff and J. A. Helayel-Neto, Phys. Rev. {\bf D67} (2003) 025016.

\bibitem{Landim}R. R. Landim, Phys. Lett. {\bf B542}160 (2002); M. A. M. Gomes
and R. R. Landim,  J.Phys. {\bf A38} (2005) 257-262.


\bibitem{sorella1}C. A. G. Sasaki, D. G. G. Sasaki and S. P. Sorella, Mod.Phys.Lett. {\bf A14} (1999) 391-396. D. G. Barci, V. E. R. Lemes, C. Linhares de Jesus, M. B. D. Silva Maia Porto, S. P. Sorella and L. C. Q. Vilar,   Nucl.Phys. {\bf B524} (1998) 765-778

\bibitem{strocchi} G. Morchio, D. Pierotti and F. Strocchi, Ann. Phys.
(N.Y.) {\bf 188} (1988) 217; A. Z. Capri and R. Ferrari, Nuovo 
Cimento {\bf A62} (1981) 273.  F. 
Strocchi, ``Selected Topics on the General Properties of 
Quantum Field Theory'', Lecture Notes in Physics, vol.51, World 
Scientific Publishing, 1993.

\bibitem{belvedereerubens} C. G. Carvalhaes, L. V. Belvedere, H. Boschi Filho and
C. P. Natividade, Ann. Phys. {\bf 258} (1997) 210; C. G. Carvalhaes, L. V. Belvedere, R. L. P. G. do Amaral and N. A.
Lemos, Ann. Phys. {\bf 269} (1998) 1.






\end{thebibliography}
\end{document}